\documentclass[prd,aps,nofootinbib,notitlepage,showpacs,preprintnumbers]
{revtex4-1}
\usepackage{graphicx,epsf,amsmath,amsfonts,amssymb,amsbsy}
\usepackage{epsfig}
\newcommand{\ds}{\displaystyle}
\newcommand{\vev}[1]{\langle#1\rangle}
\newcommand{\mat}{\left ( \begin{array}}
\newcommand{\emat}{\end{array} \right )}
\newcommand{\vect}{\left ( \begin{array}{c}}
\newcommand{\evect}{\end{array} \right )}

\begin{document}
\title{Suppression of superconductivity by inhomogeneous chiral
condensation in  the NJL$_2$ model}

\author{D. Ebert $^{1)}$, T.G. Khunjua $^{2)}$, K.G. Klimenko $^{3)}$, and V.Ch. Zhukovsky $^{2)}$}
\vspace{1cm}

 \affiliation{$^{1)}$ Institute of Physics,
Humboldt-University Berlin, 12489 Berlin, Germany}
\affiliation{$^{2)}$ Faculty of Physics, Moscow State University,
119991, Moscow, Russia} \affiliation{$^{3)}$ IHEP and University
"Dubna" (Protvino branch), 142281, Protvino, Moscow Region, Russia}

\begin{abstract}
We investigate the possibility of spatially inhomogeneous chiral and
Cooper, or superconducting, pairing in the (1+1)-dimensional model
by Chodos et al. [ Phys. Rev. D61, 045011 (2000)] generalized to
continuous chiral invariance. The consideration is performed at
nonzero values of temperature $T$ and quark number chemical potential $\mu$.
In the framework of the Fulde--Ferrel inhomogeneity
ansatz for chiral and Cooper condensates, we show that if $G_1>G_2$,
where
$G_1$ and $G_2$ are the coupling constants in the quark-antiquark
and diquark channels, then in the $(\mu,T)$-phase diagram the
superconducting phase is
suppressed by spatially inhomogeneous
chiral spiral phase with broken chiral symmetry. In contrast, in the
above mentioned original Chodos et al. model, where only the
opportunity for homogeneous condensates to be formed is taken into account, the superconducting phase is realized at sufficiently high values of $\mu$ at arbitrary values of $G_2>0$, including the interval
$0<G_2<G_1$.
\end{abstract}

\maketitle

\section{Introduction}

In the last years great attention was devoted to the investigation of dense
quark (or baryonic) matter. The interest is motivated by possible
existence of quark matter inside compact stars or its creation at
heavy ion collisions. In nature as well as in heavy ion collision
experiments, the quark matter densities are not high enough, so
consideration of its properties is not possible in the framework of
perturbation approximation of weak coupling QCD.
Usually, various effective
theories such as Nambu -- Jona-Lasinio (NJL) model, $\sigma$-model
etc. are more adequate for studying the QCD and quark matter
phase diagram in this case. As a result,  a variety
of spatially non-uniform (inhomogeneous) quark matter phases related
to chiral symmetry breaking, color superconductivity, and charged
pion condensation phenomenon etc. were predicted (see, e.g.,
\cite{3+1,nakano,nickel,maedan,Heinz:2013eu,pisarski,miransky,zfk,incera}, and references therein) at rather low values of temperature and
baryon density.

The phenomenon of spatially non-uniform quark pairing was also
intensively investigated within different (1+1)-dimensional toy
models which can mimic qualitatively the QCD phase diagram. In this
connection, it is necessary to mention the Gross-Neveu (GN) type
models that are symmetrical with respect to the discrete and
continuous chiral transformations and extended by inclusion of
baryon and isospin chemical potentials. In the framework of these
models both the inhomogeneous chiral \cite{thies,thies2,misha} and
charged pion condensation phenomena were considered
\cite{gubina,gubina2,gubina3}. (In order to overcome the
no-go theorem for spontaneous breaking of continuous
symmetry in (1+1)-dimensions,  consideration there was performed
in the large-$N$ approximation, where $N$ is the number of quark
multiplets.) Moreover, it is necessary to note that in some
quasi-one-dimensional condensed matter systems, such as
imbalanced Fermi gases \cite{Caldas}, etc., there might exist different
inhomogeneous phases.

Among a variety of GN-type models, there is one which describes
competition between quark-antiquark and diquark pairing
\cite{chodos}. In \cite{chodos}, the consideration is performed in the
supposition that chiral and superconducting condensates are
spatially homogeneous. In this case it was shown that if
$G_1>G_2$, where $G_1$ and $G_2$ are the coupling constants in the
chiral and superconducting channels, respectively, then at rather
high quark number chemical potential the superconducting phase is
realized in the system. However, the condensates in the genuine
ground state of the model may be inhomogeneous, and hence, the aim of
our paper
is to investigate such a possibility. Namely, we study the phase
structure of the extended model \cite{chodos} (which is symmetric
with respect to continuous chiral transformations), assuming that
both quark-antiquark and diquark condensates might have a spatial
inhomogeneity in the form of the Fulde--Ferrel single plane wave
ansatz \cite{ff}, for simplicity. As a result, we have shown that,
in contrast to a homogeneous ansatz for condensates,  at $G_1>G_2$
superconductivity is absent and only inhomogeneous chiral spiral (or chiral density wave)
phase occurs at arbitrary $\mu>0$ (for rather low temperature).

\section{ The model and its effective action}
\label{effaction}

Our investigation is based on a (1+1)-dimensional NJL$_2$--type
model with massless fermions, or quarks, belonging to a fundamental
multiplet of the $O(N)$ flavor group. Its Lagrangian describes the
interaction in the quark--antiquark as well as in the scalar diquark
channels:
\begin{eqnarray}
 L=\sum_{k=1}^{N}\bar \psi_k\Big [\gamma^\nu i\partial_\nu
+\mu\gamma^0\Big ]\psi_k+ \frac {G_1}N\left[\left (\sum_{k=1}^{N}\bar
\psi_k\psi_k\right )^2+\left (\sum_{k=1}^{N}\bar
\psi_k i\gamma^5\psi_k\right )^2\right ]+\frac {G_2}N\left (\sum_{k=1}^{N}
\psi_k^T\epsilon\psi_k\right )\left (\sum_{j=1}^{N}\bar
\psi_j\epsilon\bar\psi_j^T\right ), \label{1}
\end{eqnarray}
where $\mu$ is the quark number chemical potential. As it is noted
above, all fermion fields $\psi_k$ ($k=1,...,N$) form a fundamental
multiplet of $O(N)$ group. Moreover, each field $\psi_k$ is a
two-component Dirac spinor (the symbol $T$ denotes the transposition
operation). The quantities $\gamma^\nu$ ($\nu =0,1$), $\gamma^5$,
and $\epsilon$ in (1) are matrices in the spinor space,
\begin{equation}
\begin{split}
\gamma^0=\begin{pmatrix}
0&1\\
1&0\\
\end{pmatrix};\qquad
\gamma^1=\begin{pmatrix}
0&-1\\
1&0\\
\end{pmatrix}\equiv -\epsilon;\qquad
\gamma^5=\gamma^0\gamma^1=
\begin{pmatrix}
1&0\\
0&{-1}\\
\end{pmatrix}.
\end{split}\label{2}
\end{equation}
Clearly, the Lagrangian $L$ is invariant under transformations from
the internal $O(N)$ group, which is introduced here in order to make
it possible to perform all the calculations in the framework of the
nonperturbative large-$N$ expansion method. Physically more
interesting is that the model (1) is invariant under transformations
from the $U_V(1)$ quark number group: $\psi_k\to\exp
(i\alpha)\psi_k$ ($k=1,...,N$). In addition, the Lagrangian is
invariant under the continuous group $U_A(1)$ of chiral
transformations: $\psi_k\to\exp (i\alpha'\gamma^5)\psi_k$
($k=1,...,N$). \footnote{Earlier in \cite{chodos}, the similar model
symmetric under discrete $\gamma^5$ chiral transformation was
investigated. However,  only the possibility for the spatially
homogeneous chiral and diquark condensates was considered there. In
our
paper the model by Chodos et al. \cite{chodos} is generalized to the
case of continuous chiral invariance in order to study the
inhomogeneous chiral condensates in the form of chiral spirals (or
chiral density waves).} The linearized version of Lagrangian
(\ref{1}) that contains auxiliary scalar bosonic fields $\sigma
(x)$, $\pi (x)$, $\Delta(x)$, $\Delta^{*}(x)$ has the following form
\begin{eqnarray}
{\cal L}\ds =\bar\psi_k\Big [\gamma^\nu i\partial_\nu
+\mu\gamma^0 -\sigma -i\gamma^5\pi\Big ]\psi_k
 -\frac{N}{4G_1}(\sigma^2+\pi^2) -\frac N{4G_2}\Delta^{*}\Delta-
 \frac{\Delta^{*}}{2}[\psi_k^T\epsilon\psi_k]
-\frac{\Delta}{2}[\bar\psi_k \epsilon\bar\psi_k^T].
\label{3}
\end{eqnarray}
(Here and in what follows summation over repeated indices
$k=1,...,N$ is implied.) Clearly, the Lagrangians (\ref{1}) and
(\ref{3}) are equivalent, as can be seen by using the Euler-Lagrange
equations of motion for scalar bosonic fields, which take the form
\begin{eqnarray}
\sigma (x)=-2\frac {G_1}N(\bar\psi_k\psi_k),~~\pi (x)=-2\frac
{G_1}N(\bar\psi_k i\gamma^5\psi_k),~~ \Delta(x)=-2\frac
{G_2}N(\psi_k^T\epsilon\psi_k),~~ \Delta^{*}(x)=-2\frac
{G_2}N(\bar\psi_k \epsilon\bar\psi_k^T). \label{4}
\end{eqnarray}
One can easily see from (\ref{4}) that the (neutral) fields
$\sigma(x)$ and $\pi(x)$ are real quantities, i.e.
$(\sigma(x))^\dagger=\sigma(x)$, $(\pi(x))^\dagger=\pi(x)$ (the
superscript symbol $\dagger$ denotes the Hermitian conjugation), but
the (charged) diquark scalar fields $\Delta(x)$ and $\Delta^*(x)$
are mutually Hermitian conjugated complex quantities, so that
$(\Delta(x))^\dagger= \Delta^{*}(x)$ and vice versa. Clearly, all
the fields (\ref{4}) are singlets with respect to the $O(N)$ group.
\footnote{Note that the $\Delta (x)$ field is a flavor O(N) singlet,
since the representations of this group are real.} If the scalar
diquark field $\Delta(x)$ has a nonzero ground state expectation
value, i.e.\  $\vev{\Delta(x)}\ne 0$, the Abelian quark number
$U_V(1)$ symmetry of the model is spontaneously broken down.
However, if $\vev{\sigma (x)}\ne 0$ then the continuous $U_A(1)$
chiral symmetry of the model is spontaneously broken.

Let us now study the phase structure of the four-fermion model (1)
by starting with the equivalent semi-bosonized Lagrangian (\ref{3}).
In the leading order of the large-$N$ approximation, the effective
action ${\cal S}_{\rm {eff}}(\sigma,\pi,\Delta,\Delta^{*})$ of the
considered model is expressed in terms of the path integral over
fermion fields:
$$
\exp(i {\cal S}_{\rm {eff}}(\sigma,\pi,\Delta,\Delta^{*}))=
  \int\prod_{l=1}^{N}[d\bar\psi_l][d\psi_l]\exp\Bigl(i\int {\cal
  L}\,d^2 x\Bigr),
$$
where
\begin{eqnarray}
&&{\cal S}_{\rm {eff}} (\sigma,\pi,\Delta,\Delta^{*}) =-\int
d^2x\left [\frac{N}{4G_1}(\sigma^2(x)+\pi^2(x))+
\frac{N}{4G_2}\Delta (x)\Delta^{*}(x)\right ]+ \widetilde {\cal
S}_{\rm {eff}}. \label{5}
\end{eqnarray}
The fermion contribution to the effective action, i.e.\  the term
$\widetilde {\cal S}_{\rm {eff}}$ in (\ref{5}), is given by:
\begin{equation}
\exp(i\widetilde {\cal S}_{\rm
{eff}})=\int\prod_{l=1}^{N}[d\bar\psi_l][d\psi_l]\exp\Bigl\{i\int\Big
[\bar\psi_k(\gamma^\nu i\partial_\nu +\mu\gamma^0 -\sigma
-i\gamma^5\pi)\psi_k -
\frac{\Delta^{*}}{2}(\psi_k^T\epsilon\psi_k)
-\frac{\Delta}{2}(\bar\psi_k \epsilon\bar\psi_k^T)\Big ]d^2
x\Bigr\}. \label{6}
\end{equation}
The ground state expectation values $\vev{\sigma(x)}$,
$\vev{\pi(x)}$, etc of the composite bosonic fields are determined
by the saddle point equations,
\begin{eqnarray}
\frac{\delta {\cal S}_{\rm {eff}}}{\delta\sigma (x)}=0,~~~~~
\frac{\delta {\cal S}_{\rm {eff}}}{\delta\pi (x)}=0,~~~~~
\frac{\delta {\cal S}_{\rm {eff}}}{\delta\Delta (x)}=0,~~~~~
\frac{\delta {\cal S}_{\rm {eff}}}{\delta\Delta^* (x)}=0. \label{7}
\end{eqnarray}
In vacuum, i.e. in the state corresponding to an empty space with
zero particle density and zero value of the chemical potential
$\mu$, the above mentioned quantities $\vev{\sigma(x)}$, etc.
(\ref{7}) do not depend on space coordinates. However, in a dense
medium, when $\mu\ne 0$, the ground state expectation values of
bosonic fields (\ref{4}) might have a nontrivial dependence on the
spatial coordinate $x$. For simplicity, in this paper we will use the
following well-known ansatz:
\begin{eqnarray}
\vev{\sigma(x)}=M\cos (2bx),~~~\vev{\pi(x)}=M\sin
(2bx),~~~\vev{\Delta(x)}=\Delta\exp(-2ib'x),~~~
\vev{\Delta^*(x)}=\Delta\exp(2ib'x), \label{8}
\end{eqnarray}
where $M,b,b'$ and $\Delta$ are real constant quantities. (It means
that we suppose for $\vev{\sigma(x)}$ and $\vev{\pi(x)}$ the chiral
spiral (or chiral density wave) ansatz, and the Fulde--Ferrel
\cite{ff} single plane wave one for diquark condensates.) In fact,
$M,b,b'$ and $\Delta$ are coordinates of the global minimum point of
the thermodynamic potential (TDP) $\Omega (M,b,b',\Delta)$.
\footnote{Here and in what follows we will use a rather
conventional notation "global" minimum in the sense that among all
our numerically found local minima the thermodynamical potential
takes in their case the lowest value. This does not exclude the
possibility that there exist other inhomogeneous condensates,
different from (\ref{8}), which lead to ground states with even
lower values of the TDP.} In the leading order of the large-$N$ expansion it is defined by the following expression:
\begin{equation*}
\int d^2x \Omega (M,b,b',\Delta)=-\frac{1}{N}{\cal S}_{\rm
{eff}}\{\sigma(x),\pi(x),\Delta (x),\Delta^*(x)\}\Big|_{\sigma
    (x)=\vev{\sigma(x)},\pi(x)=\vev{\pi_a(x)},...} ,
\end{equation*}
which gives
\begin{eqnarray}
\int d^2x\Omega (M,b,b',\Delta)\,\,&=&\,\,\int d^2x\left
(\frac{M^2}{4G_1}+\frac{\Delta^2}{4G_2}\right )+\frac{i}{N}\ln\left
( \int\prod_{l=1}^{N}[d\bar\psi_l][d\psi_l]\exp\Big (i\int d^2 x\Big
[\bar\psi_k {\cal D}\psi_k\right.\nonumber\\&& \left.-
\frac{\Delta\exp(2ib'x)}{2}(\psi_k^T\epsilon\psi_k)
-\frac{\Delta\exp(-2ib'x)}{2}(\bar\psi_k \epsilon\bar\psi_k^T)\Big ]
\Big )\right ), \label{9}
\end{eqnarray}
where ${\cal D}=\gamma^\rho i\partial_\rho +\mu\gamma^0
-M\exp(2i\gamma^5bx)$. To proceed, let us introduce in (\ref{9}) the
new fermion fields, $q_k=\exp[i(b'+\gamma^5b)x]\psi_k$ and $\bar
q_k= \bar\psi_k \exp[i(\gamma^5b-b')x]$. Since this transformation
of fermion fields does not change the path integral measure in
(\ref{9})  \footnote{This nontrivial fact follows from the
investigations by Fujikawa \cite{fujikawa}, who established that a
chiral transformation of spinor fields changes the path integral
measure only in the case, when there is  interaction between spinor
and gauge fields.},
the expression (\ref{9}) for the thermodynamic
potential is easily transformed into the following one:
\begin{eqnarray}
\int d^2x\Omega (M,b,b',\Delta)&=&\int d^2x\left
(\frac{M^2}{4G_1}+\frac{\Delta^2}{4G_2}\right ) \nonumber\\
&+&\frac{i}{N}\ln\left ( \int\prod_{l=1}^{N}[d\bar q_l][d
q_l]\exp\Big (i\int d^2 x\Big
[\bar q_k  D q_k
-\frac{\Delta}{2}(q_k^T\epsilon q_k) -\frac{\Delta}{2}(\bar q_k
\epsilon\bar q_k^T)\Big ] \Big )\right ) \label{10},
\end{eqnarray}
where
\begin{equation}
D=\gamma^\nu i\partial_\nu +(\mu-b)\gamma^0-M+\gamma^1b'.
\label{110}
\end{equation}
The path integration in this expression can be evaluated (see, e.g.,
Appendix B of the paper \cite{kzz}), so we have for the TDP
\begin{eqnarray}
\Omega (M,b,b',\Delta)\equiv\Omega^{un} (M,b,b',\Delta)&=&
\frac{M^2}{4G_1}+\frac{\Delta^2}{4G_2}
+\frac{i}{2}\int\frac{d^2p}{(2\pi)^2}\ln\Big (\lambda_1(p)\lambda_2(p)\Big ), \label{11}
\end{eqnarray}
where ($\tilde\mu=\mu-b$)
\begin{eqnarray}
\lambda_{1,2}(p)=p_0^2-\tilde\mu^2-p_1^2+b'^2+M^2-\Delta^2\pm
2\sqrt{M^2 p_0^2-M^2 p_1^2+\tilde\mu^2 p_1^2-2 p_0 b'\tilde\mu
p_1+p_0^2 b'^2}\label{A8}
\end{eqnarray}
and superscription ``un'' denotes the unrenormalized quantity. In
the following we will study the behavior of the global minimum point
of this TDP as a function of dynamical variables $M,b,b',\Delta$ vs
the external parameter $\mu$ in two qualitatively different cases:
i) The case of homogeneous condensates, i.e. when in (\ref{11}) both
$b$ and $b'$ are supposed to be put to zero by hand, i.e. without any
proof, ii) The case of spatially inhomogeneous condensates, i.e.
when the quantities $b$ and $b'$ are defined dynamically by the gap
equations of the TDP (\ref{11}). Note finally that the expression
(\ref{11}) is the TDP of the initial system at zero temperature $T$.
The consideration of the case $T\ne 0$ will also be taken into
account in the subsequent sections.

\section{Homogeneous ansatz for both condensates: $b=0$, $b'=0$}
\label{hom}

In this section we assume both chiral and
superconducting condensates to be homogeneous, i.e. putting $b=0$ and $b'=0$.
Then the TDP (\ref{11}) is reduced to the following expression
\begin{eqnarray}
\Omega^{un} (M,\Delta)= \frac{M^2}{4G_1}+\frac{\Delta^2}{4G_2}
+\frac{i}{2}\int\frac{d^2p}{(2\pi)^2}\ln\Big
[(p_0^2-E_+^2)(p_0^2-E_-^2)\Big ],  \label{12}
\end{eqnarray}
where (note, $E_-^2$ is a nonnegative quantity)
\begin{eqnarray}
E_\pm^2=\mu^2+p_1^2+M^2+\Delta^2\pm 2\sqrt{M^2\Delta^2+\mu^2(p_1^2+M^2)}.
 \label{13}
\end{eqnarray}
Integrating in (\ref{12}) over $p_0$ (see Ref. \cite{Ebert:2009ty}
for  similar integrals), one obtains for the TDP
$\Omega^{un}(M,\Delta)$:
\begin{eqnarray}
\Omega^{un} (M,\Delta)&=&
\frac{M^2}{4G_1}+\frac{\Delta^2}{4G_2}-
\int_{-\infty}^\infty\frac{dp_1}{4\pi}\Big (E_++E_-\Big ). \label{14}
\end{eqnarray}
Clearly, without loss of generality one can study the TDP
(\ref{12})-(\ref{14})  in the region $M\ge 0$ and $\Delta\ge 0$ at
$\mu\ge 0$.

\subsection{The renormalization}

Formally, the TDP (\ref{12})-(\ref{14}) is an ultraviolet divergent
quantity. To renormalize it, i.e. to obtain a finite expression for
it, we first need to regularize the TDP (\ref{14}) by cutting off its
integration region, $|p_1|<\Lambda$:
\begin{eqnarray}
\Omega^{un} (M,\Delta)\to\Omega^{reg} (M,\Delta)&=&
\frac{M^2}{4G_1}+\frac{\Delta^2}{4G_2}-
\int_{0}^\Lambda\frac{dp_1}{2\pi}\Big (E_++E_-\Big ). \label{15}
\end{eqnarray}
Second, we must find such dependencies of the bare coupling
constants $G_1\equiv G_1(\Lambda)$ and $G_2\equiv G_2(\Lambda)$ vs
$\Lambda$ that in the limit $\Lambda\to\infty$ one can obtain from
$\Omega^{reg} (M,\Delta)$ a finite expression. To get the quantities
$G_1(\Lambda)$ and $G_2(\Lambda)$ let us first use in (\ref{15}) the
following asymptotic expansion at $p_1\to\infty$:
 \begin{eqnarray}
E_++E_-=2E+\frac{\Delta^2}{E}+{\cal O}(1/E^2),\label{16}
\end{eqnarray}
where $E=\sqrt{M^2+p_1^2}$. Then, substituting (\ref{16}) into
(\ref{15}) and integrating there over $p_1$, we obtain the following
asymptotic expansion of the regularized TDP at $\Lambda\to\infty$:
 \begin{eqnarray}
\Omega^{reg} (M,\Delta)=\frac{M^2}{4G_1}+\frac{\Delta^2}{4G_2}
-\frac{1}{2\pi}\Big (\Lambda^2+(M^2+\Delta^2)\ln\frac{\Lambda+\sqrt{\Lambda^2+M^2}}{M}\Big )+{\cal O}(\Lambda^0).\label{17}
\end{eqnarray}
Now it is obvious from (\ref{17}) that all ultraviolet divergences (up
to unessential term -$\Lambda^2/2\pi$) of the thermodynamic potential
(\ref{15}) can be removed, if
 \begin{eqnarray}
\frac{1}{4G_1}\equiv\frac{1}{4G_1(\Lambda)}=\frac{1}{2\pi}\ln\frac{2\Lambda}{M_1},~~~\frac{1}{4G_2}\equiv\frac{1}{4G_2(\Lambda)}=\frac{1}{2\pi}\ln\frac{2\Lambda}{M_2},\label{18}
\end{eqnarray}
where $M_1$ and $M_2$ are some finite and cutoff independent
parameters of the model. In addition, $M_1$ and $M_2$ are also renormalization invariant,
i.e. they do not depend on normalization points. (The physical
meaning of these parameters will be discussed below.)
Hence, in the limit $\Lambda\to \infty$ one can obtain from (\ref{15})
a finite renormalization invariant expression for the TDP
\begin{eqnarray}
\Omega(M,\Delta)&=&\lim_{\Lambda\to\infty}\left\{\Omega^{reg}
(M,\Delta)\Big |_{G_1= G_1(\Lambda),G_2= G_2(\Lambda)}+\frac{\Lambda^2}{2\pi}\right\}.\label{19}
\end{eqnarray}
In the following, instead of treating the results in terms of two
dimensional parameters $M_{1,2}$, we will use, as in the paper
\cite{chodos}, one dimensional, $M_1$, and one dimensionless,
$\delta$, parameters, i.e.
 \begin{eqnarray}
\frac{\delta}{4\pi}\equiv\frac{1}{4G_2}-\frac{1}{4G_1}=\frac{1}{2\pi}\ln\frac{M_1}{M_2}.
\label{20}
\end{eqnarray}
Since $M_1$ and $\delta$ might be considered as  free model
parameters, it is clear
that the renormalization procedure of the NJL$_2$ model (1) is
accompanied by the {\it partial dimensional
transmutation phenomenon}. Indeed, in the initial unrenormalized
expression (\ref{12}) for $\Omega^{un}(M,\Delta)$ two dimensionless
bare coupling constants $G_{1,2}$ are present, whereas after
renormalization the thermodynamic potential (\ref{19}) is
characterized in our choice of a parameterization by one dimensional,
$M_1$, and one dimensionless, $\delta$, free model parameters.

\subsection{The phase structure}

In this subsection we will study the phase structure of the model
(1) in three cases, first at $\mu=0$ and $T=0$ (it is the vacuum
case), second at $\mu>0$, $T=0$ and, finally, at $\mu>0$ and $T>0$.
All the condensates are still supposed to be homogeneous.

{\bf The vacuum case ($\mu=0$, $T=0$).} Putting $\mu=0$ in
(\ref{15}), we have for the regularized effective potential
$V_0^{reg}(M,\Delta)$ ( TDP in vacuum is usually called  effective
potential) the following expression:
\begin{eqnarray}
V_0^{reg} (M,\Delta)&=&
\frac{M^2}{4G_1}+\frac{\Delta^2}{4G_2}-
\int_{0}^\Lambda\frac{dp_1}{2\pi}\Big (\sqrt{p_1^2+(M+\Delta)^2}+\sqrt{p_1^2+(M-\Delta)^2}\Big ). \label{21}
\end{eqnarray}
After integration in (\ref{21}) over $p_1$ we should renormalize the
obtained expression, i.e. to put $G_1=G_1(\Lambda)$,
$G_2=G_2(\Lambda)$ (see in (\ref{18})) and then to find a limit at
$\Lambda\to\infty$ of the expression (\ref{19}). In terms of $M_1$
and $\delta$ the renormalized effective potential looks like
\begin{eqnarray}
4\pi V_0(M,\Delta)&=&\delta\Delta^2-\Delta^2-M^2+(M-\Delta)^2\ln\left |\frac{M-\Delta}{M_1}\right |+(M+\Delta)^2\ln\left (\frac{M+\Delta}{M_1}\right ).
 \label{22}
\end{eqnarray}
If $\delta>0$, i.e., as is easily seen from (\ref{18}) and
(\ref{20}), at $G_1>G_2$, the global minimum of the effective
potential (\ref{22}) lies at the point $(M=M_1,\Delta=0)$. This
means that if  interaction in the quark-antiquark channel is
greater than in the diquark one, then the chiral symmetry $U_A(1)$
of the model is spontaneously broken down and fermions acquire
dynamically a nonzero mass, which is equal just to the free model
parameter $M_1$. However, if $\delta<0$, i.e. at $G_1<G_2$, then the
global minimum  of the effective potential (\ref{22}) is arranged at
the point $(M=0,\Delta=\Delta_0(\delta))$, where
$\Delta_0(\delta)=M_1\exp(-\delta/2)=M_2$. Since in this case only
the diquark condensate is nonzero, the fermion number $U_V(1)$
symmetry is spontaneously broken down.

{\bf The case $\mu>0$ and $T=0$.}
\label{mu}
Obviously, at $\mu>0$ the regularized TDP (\ref{15}) can be presented in the form
\begin{eqnarray}
\Omega^{reg} (M,\Delta)&=&V_0^{reg} (M,\Delta)
-\int_{0}^\Lambda\frac{dp_1}{2\pi}\Big (E_++E_--\sqrt{p_1^2+(M+\Delta)^2}-\sqrt{p_1^2+(M-\Delta)^2}\Big ), \label{23}
\end{eqnarray}
where $V_0^{reg} (M,\Delta)$ is given in (\ref{21}). As a result, at
$\Lambda\to\infty$ one can obtain from (\ref{23}) and (\ref{19}) the
renormalized TDP
\begin{eqnarray}
\Omega (M,\Delta)&=&V_0 (M,\Delta)
-\int_{0}^\infty\frac{dp_1}{2\pi}\Big (E_++E_--\sqrt{p_1^2+(M+\Delta)^2}-\sqrt{p_1^2+(M-\Delta)^2}\Big ), \label{24}
\end{eqnarray}
where $V_0 (M,\Delta)$ is presented in (\ref{22}). After numerical
investigations of the function (\ref{24}), it is clear that its
global minimum might lie only in the points of the form $(M\ge
0,\Delta=0)$ or $(M=0,\Delta\ge 0)$. As a result, for further
consideration it is enough to reduce the TDP (\ref{24}) to the $M$-
and $\Delta$-axes, where it looks like
\begin{eqnarray}
4\pi\Omega (M,\Delta=0)&=&M^2\Big (\ln\frac{M^2}{M_1^2}-1\Big )+2\theta (\mu-M)\Big [M^2\ln\frac{\mu+\sqrt{\mu^2-M^2}}{M}-\mu\sqrt{\mu^2-M^2}\Big ], \label{37}\\
4\pi\Omega (M=0,\Delta)&=&\delta\Delta^2+\Delta^2\Big (\ln\frac{\Delta^2}{M_1^2}-1\Big )-2\mu^2, \label{38}
\end{eqnarray}
respectively. Apart from a trivial extremum at the origin for
both functions (\ref{37}) and (\ref{38}), the TDP (\ref{37}) has a
nontrivial extremum at the point $M=M_1$, whereas the second TDP,
i.e. the function (\ref{38}), has a nontrivial minimum point at
$\Delta=M_1\exp (-\delta/2)=M_2$. Comparing the values of the
functions (\ref{37})-(\ref{38}) at these extrema, we obtain the
following evolution of the global minimum point (GMP) of the TDP
(\ref{24}) vs $\mu$ and, as a result, the phase portrait of the
initial model.

First, let us suppose that $\delta>0$, i.e. $G_1>G_2$. Then at
sufficiently low $\mu$-values the GMP of the TDP (\ref{24}) lies at
the point $(M=M_1,\Delta=0)$, which corresponds to the chiral
symmetry breaking phase of the model. In contrast, at sufficiently
high values of the chemical potential the GMP of the TDP (\ref{24})
is arranged at the point $(M=0,\Delta=M_1\exp (-\delta/2))$. In this
case the diquark condensation, or in other words, superconducting,
phase is
realized. The critical value $\mu_c$, at which the first order
phase transition between these phases takes pace, can be easily calculated,
\begin{eqnarray}
\mu_c=M_1\sqrt{\frac{1-e^{-\delta}}{2}}~~.\label{39}
\end{eqnarray}
It follows from (\ref{39}) that at $\delta\to\infty$ we have
$\mu_c\to M_1/\sqrt{2}$, i.e. even at arbitrary small interaction in
the diquark channel there is a superconducting phase in the model at
$\mu>\mu_c$. It is the so-called Cooper instability of the model.

 Second, if $\delta<0$, then we have found only the
diquark condensation phase for arbitrary values of $\mu$. It means
that in the case, when an interaction in the diquark channel is
stronger than in the quark--antiquark one the quark number $U_V(1)$
symmetry is spontaneously broken down and superconducting phase is
formed in the model at $\mu>0$.

Finally, few words about the particle  density $n=-\partial\Omega
(M_0,\Delta_0)/\partial\mu$ of the system at zero temperature (where
$(M_0,\Delta_0)$ is the GMP of the TDP (\ref{24})). Suppose that
$\delta>0$ and $\mu<\mu_c$ (\ref{39}), i.e., that we are in the
chiral symmetry breaking phase. Since in this case $M_0=M_1$ and
$\Delta_0=0$, one can easily obtain from the expression (\ref{37})
that $n\equiv 0$ in the chirally broken phase. However, at
$\mu>\mu_c$ the system is in the superconducting phase, where
$M_0=0$ and $\Delta_0=M_1\exp (-\delta/2))$. As a result, it follows
from the expression (\ref{38}) that in this phase the particle
density is nonzero, $n=\mu/\pi$. The last expression for particle
density is also valid for superconducting phase at $\delta<0$.

{\bf The case $T\ne 0$.} In order to include  temperature into our
consideration, let us start from the unrenormalized expression
(\ref{12}) for the TDP, where one should perform the following
standard replacements:
\begin{eqnarray}
\int_{-\infty}^{\infty}\frac{dp_0}{2\pi}\big (\cdots\big )\to
iT\sum_{n=-\infty}^{\infty}\big (\cdots\big ),~~~~p_{0}\to
p_{0n}\equiv i\omega_n \equiv i\pi T(2n+1),~~~n=0,\pm 1, \pm 2,...,
\label{430}
\end{eqnarray}
i.e. the $p_0$-integration should be replaced by the summation over
an infinite set of Matsubara frequencies $\omega_n$. Summing over
Matsubara frequencies in the obtained expression  (the corresponding
technique is presented, e.g., in \cite{jacobs}), one can find for
the unrenormalized temperature dependent TDP $\Omega^{un}_{\scriptscriptstyle{T}}(M,\Delta)$ the following expression:
\begin{eqnarray}
\Omega^{un}_{\scriptscriptstyle{T}}(M,\Delta)\!
&=&\frac{M^2}{4G_1}+\frac{\Delta^2}{4G_2}-\int_{0}^{\infty}\frac{dp_1}{2\pi}
\Big\{E_++E_-\Big\}-T\int_{0}^{\infty}\frac{dp_1}{\pi}\ln\big\{ [1+e^{-\beta
E_+}][1+e^{-\beta
E_-}]\big\}\nonumber\\&=&\Omega^{un}(M,\Delta)-T\int_{0}^{\infty}\frac{dp_1}{\pi}\ln\big\{ [1+e^{-\beta
E_+}][1+e^{-\beta E_-}]\big\}, \label{1201}
\end{eqnarray}
where $\beta=1/T$, $E_\pm$ are given in (\ref{13}) and
$\Omega^{un}(M,\Delta)$ is just the unrenormalized TDP (\ref{14}) at
$T=0$. Since the last integral in (\ref{1201}) is convergent, in
order to obtain a finite renormalized TDP at nonzero temperature, we
should simply renormalize the TDP $\Omega^{un}(M,\Delta)$, as it was
done in the previous subsection \ref{mu}. Thus, in the case of
homogeneous condensates we obtain the following renormalized $T$-
and $\mu$-dependent TDP:
\begin{eqnarray}
\Omega_{\scriptscriptstyle{T}}(M,\Delta)\!
&=&\Omega(M,\Delta)-T\int_{0}^{\infty}\frac{dp_1}{\pi}\ln\big\{ [1+e^{-\beta
E_+}][1+e^{-\beta E_-}]\big\}, \label{120}
\end{eqnarray}
where $\Omega(M,\Delta)$ is the renormalized TDP (\ref{24}) at zero
temperature and $\mu>0$. For particular values of $\delta=1$ and
$\delta=-1$ the numerical investigations lead to the
$(\mu,T)$--phase portraits shown in Fig. 1 and Fig. 2,
respectively. Note that in Fig. 1, there is a phase transition of
the first order on the boundary between the chiral symmetry breaking
phase (in which $\Delta=0$, $M\ne 0$) and diquark, or superconducting,
phase with $M=0$, $\Delta\ne 0$. However, on the boundary of the
symmetrical phase there are second-order phase transitions in both
figures.
\begin{figure}
\includegraphics[width=0.45\textwidth]{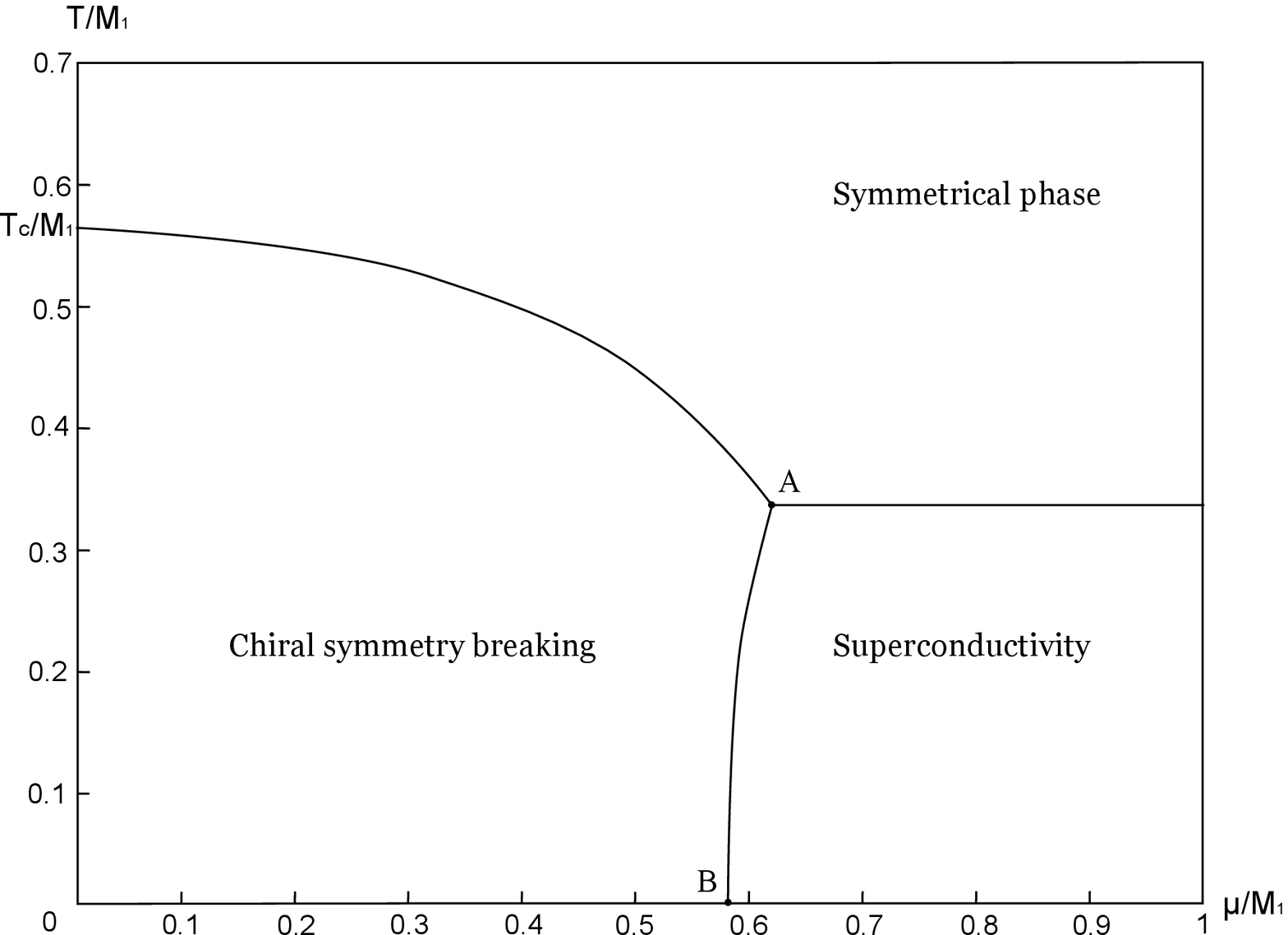}
\hfill
\includegraphics[width=0.45\textwidth]{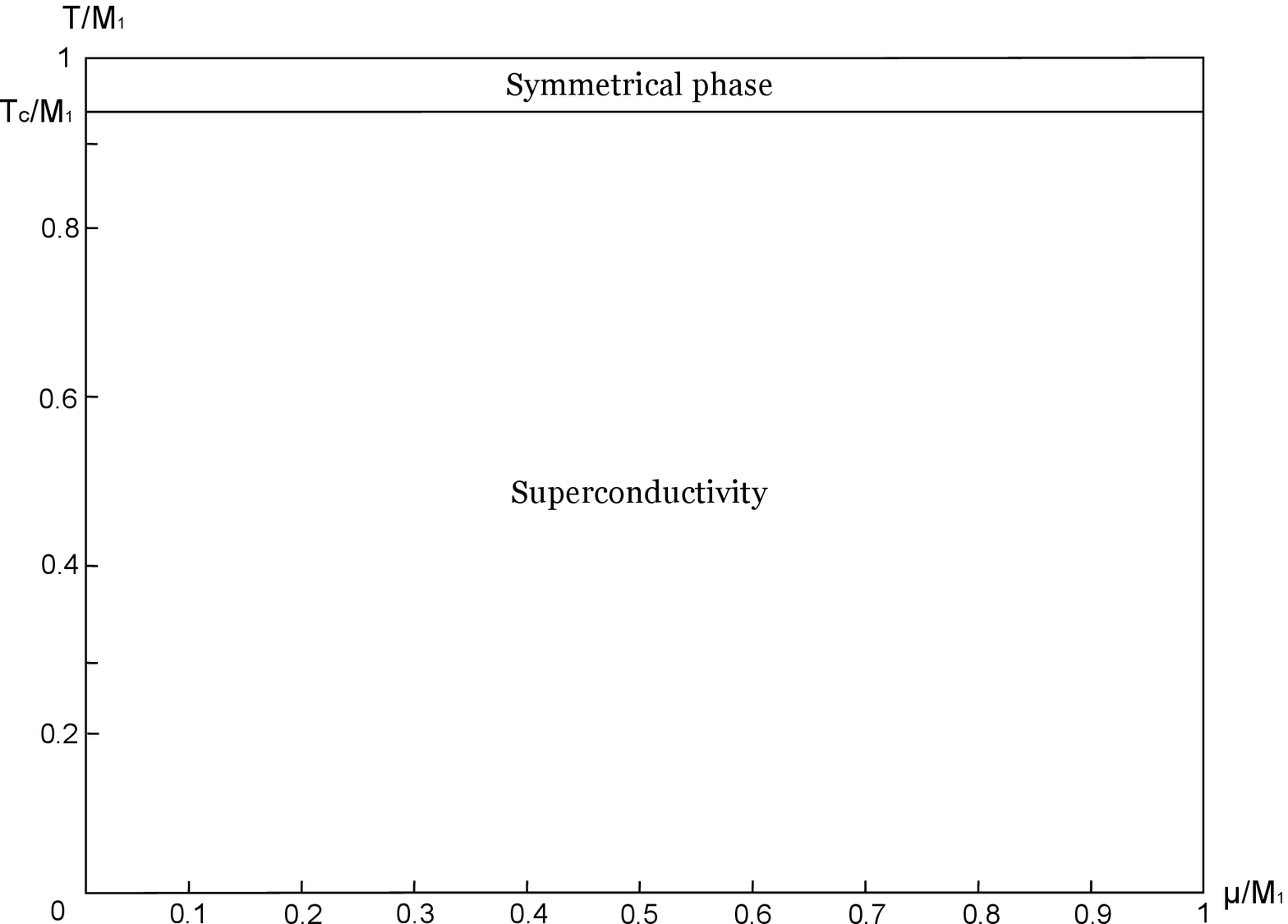}
\\
\parbox[t]{0.45\textwidth}{
 \caption{The $(\mu,T)$--phase structure of the model at $\delta=1$
for the homogeneous case of the ansatz (\ref{8}) for condensates
$(b=b'=0)$. Here $T_c\approx 0.57M_1$. The $(\mu,T)$-coordinates of
the points A and B are the following: A$\equiv$A$(0.62M_1,0.33M_1)$,
B$\equiv$B$(0.59M_1,0)$. In the symmetrical phase $M=\Delta=0$.} } \hfill
\parbox[t]{0.45\textwidth}{
 \caption{The $(\mu,T)$--phase structure of the model at $\delta=-1$
both for homogeneous $(b=b'=0)$ and inhomogeneous $(b\ne 0,b'\ne 0)$ cases of the ansatz (\ref{8}) for condensates. Here $T_c\approx 0.93M_1$.  In the symmetrical phase $M=\Delta=b=b'=0$. }}
\end{figure}
It is interesting also to look at the behavior of the particle
density
\begin{eqnarray}
n=-\partial\Omega_{\scriptscriptstyle{T}} (M_0,\Delta_0)/\partial\mu
 \label{12030}
\end{eqnarray}
of the system at nonzero temperature. In this formula $M_0$ and
$\Delta_0$ stand for coordinates of the GMP of the TDP (\ref{120}).
(It is evident that $M_0$ and $\Delta_0$ are temperature dependent
quantities.) Hence, in the chiral symmetry breaking phase, where
$M_0\ne 0$ but $\Delta_0=0$, the particle density (\ref{12030}),
denoted in this phase as $n_{\scriptscriptstyle{CSB}}$, looks like
\begin{eqnarray}
n_{\scriptscriptstyle{CSB}}(\mu,T)&=&\frac{\theta(\mu-M_0)}{\pi}\sqrt{\mu^2-M_0^2}+\frac{1}{\pi}\int_{0}^{\infty}dp_1\left\{\frac{\theta(E-\mu)}{1+e^{\beta (E-\mu)}}
-\frac{\theta(\mu-E)}{1+e^{\beta (\mu-E)}}-\frac{1}{1+e^{\beta (E+\mu)}}\right\},
 \label{1203}
\end{eqnarray}
where $E=\sqrt{p_1^2+M_0^2}$. Assuming in (\ref{1203}) that $M_0=0$,
one can obtain the particle density ($\equiv
n_{\scriptscriptstyle{SYM}}$) in the symmetric phase of the model
\begin{eqnarray}
n_{\scriptscriptstyle{SYM}}(\mu,T)&=&\frac{\mu}{\pi}+\frac{1}{\pi}\int_{0}^{\infty}dp_1\left\{\frac{\theta(p_1-\mu)}{1+e^{\beta (p_1-\mu)}}
-\frac{\theta(\mu-p_1)}{1+e^{\beta (\mu-p_1)}}-\frac{1}{1+e^{\beta (p_1+\mu)}}\right\}\equiv\frac{\mu}{\pi},
 \label{1204}
\end{eqnarray}
i.e. in the symmetrical phase of the model the particle density does
not depend on temperature.
Finally, since in the superconducting phase $M_0= 0$ and $\Delta_0\ne
0$, it is possible to obtain, after some manipulations, the following
expression for the particle density ($\equiv
n_{\scriptscriptstyle{SC}}$) in this phase:
\begin{eqnarray}
n_{\scriptscriptstyle{SC}}(\mu,T)&=&\frac{\mu}{\pi}-\frac{1}{\pi}\int_{0}^{\infty}dp_1\left\{\frac{\mu+p_1}{\epsilon_+ [1+e^{\beta\epsilon_+}]}+\frac{\mu-p_1}{\epsilon_- [1+e^{\beta\epsilon_-}]}\right\}\equiv\frac{\mu}{\pi}.
 \label{1205}
\end{eqnarray}
In (\ref{1205}) we use the notations
$\epsilon_\pm=\sqrt{\Delta_0^2+(\mu\pm p_1)^2}$. It is clear that in
the superconducting phase the particle density is also a temperature
independent quantity.

It is evident that the plots of the functions
$n_{\scriptscriptstyle{CSB}}(\mu,T)$,
$n_{\scriptscriptstyle{SYM}}(\mu,T)$ and
$n_{\scriptscriptstyle{SC}}(\mu,T)$ are some surfaces in three
dimensional parametric space $(\mu,T,n)$. As a whole, the
combination of these surfaces are no more than the plot of the
particle density (\ref{12030}) $n(\mu,T)$ vs $\mu,T$. At $\delta =1$
it is depicted in Fig. 3. Recall, the $(\mu,T)$-phase portrait of
the model was already presented in Fig. 1 at $\delta=1$. However,
sometimes the phase diagram in terms of the fermion number density
$n$ and temperature $T$ is more informative. To obtain the
$(n,T)$-phase portrait of the model one should simply  construct
projections of the boundaries of the surfaces
$n_{\scriptscriptstyle{CSB}}(\mu,T)$,
$n_{\scriptscriptstyle{SYM}}(\mu,T)$ and
$n_{\scriptscriptstyle{SC}}(\mu,T)$ (in Fig. 3 these boundaries are
represented as thick solid lines) onto the $(n,T)$-coordinate plane.
As a result, we will divide the $(n,T)$-plane into several different
phases. \footnote{It is evident that one reproduces the
$(\mu,T)$-phase portrait of Fig. 1, when finds the projections of
all boundaries of the above mentioned surfaces onto the
$(\mu,T)$-coordinate plane of Fig. 3.} Performing this procedure in
the case $\delta =1$ (see Fig. 3), it is possible to find the
$(n,T)$-phase diagram at $\delta =1$. It is presented in Fig. 4,
where you can see the usual (pure) chiral symmetry breaking,
superconducting and symmetrical phases of the model. In addition,
there is a region of the figure corresponding to a co-existence (or
mixture) of the chiral symmetry breaking and superconducting phases.
(The co-existence of two arbitrary phases means that in the space
filled by one of them there are bubbles of another phase and vice
versa.) In this case in the chiral symmetry breaking phase there
might appear at some temperature a bubble of a more dense
superconducting phase. If energy is provided to the system, then (at
fixed value of a chemical potential) the size of the bubble is
increased, i.e. the average particle density of the system is also
increased. The process can be presented in Fig. 4 as a movement
along the straight line parallel to the density axis, which crosses
the region of the phase co-existence. Note also that the region of
phase mixture in Fig. 4 is represented in Fig. 1 as a first-order
phase transition curve AB (in Fig. 3 it corresponds to a vertical
cylindrical surface, connecting the $n_{\scriptscriptstyle{CSB}}$
and $n_{\scriptscriptstyle{SC}}$ surfaces).

The above mentioned way for constructing $(n,T)$-phase diagram of the
model can be used in the case $\delta<0$ as well. Qualitatively, it is
the same as the phase portrait of Fig. 2, in which one should simply
rename the horizontal $\mu$-axis in favor of $n$-axis.

It is clear from Figs 1, 2 and 4 that in the case of homogeneous
condensates the temperature of a chiral symmetry restoration phase
transition depends strongly on values of chemical potential $\mu$
and/or particle density $n$. In contrast, the critical temperature
$T_{crit}$ of a transition between symmetrical and superconducting
phases does not depend on  $\mu$ and/or $n$. However, in the
framework of the model under consideration this property is valid
only in the (1+1)-dimensional spacetime. In higher dimensions (see,
e.g., the paper \cite{kzz}, where just the same model (1), but in
the (2+1)-dimensional spacetime, was investigated) the critical
temperature of a transition between symmetrical and superconducting
phases is already a $\mu$-dependent quantity. The conclusion is also
supported by results of the paper \cite{Ebert:2010eq}, where the
phase structure of a (3+1)-dimensional version of the model (1) was
considered. In \cite{Ebert:2010eq}, one coordinate was compactified
($L$ is the radius of a compactification). It was shown there that
in the case of antiperiodic boundary conditions for fermion fields
the critical value of $1/L$, at which superconductivity is
transformed into symmetric phase, depends on $\mu$ (see Fig. 9 of
\cite{Ebert:2010eq}). Since at  antiperiodic boundary conditions the
quantity $1/L$ behaves in many respects like temperature, one can
believe that in the framework of the (3+1)-dimensional model (1)
$T_{crit}$ also depends on $\mu$.
\begin{figure}
\includegraphics[width=0.6\textwidth]{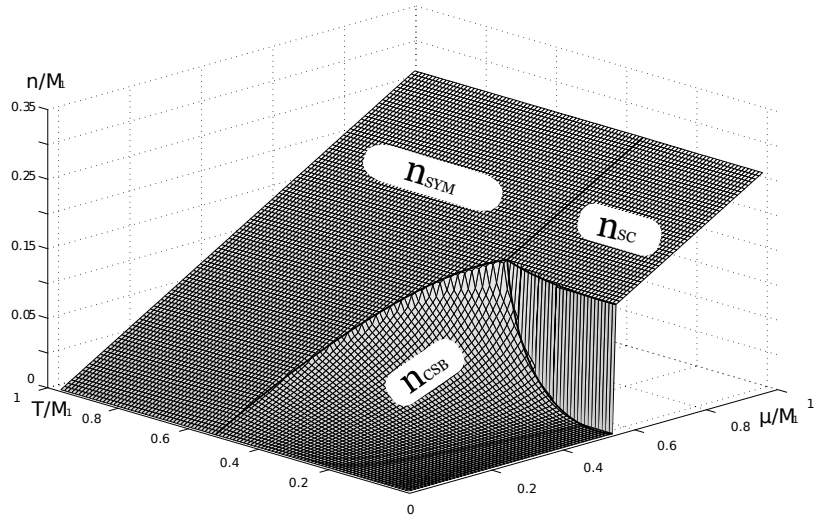}
\\
 \caption{The particle density $n$ vs $(\mu,T)$ at $\delta=1$ in the
 homogeneous case of the ansatz (\ref{8}) for condensates $(b=b'=0)$. The surfaces $n_{\scriptscriptstyle{CSB}}$,
$n_{\scriptscriptstyle{SYM}}$ and
$n_{\scriptscriptstyle{SC}}$ represent the behavior of the particle
density $n(\mu,T)$ (\ref{12030}) in the chiral symmetry breaking-,
symmetrical- and superconducting phases of the model,
respectively.}
\end{figure}
\begin{figure}
\includegraphics[width=0.45\textwidth]{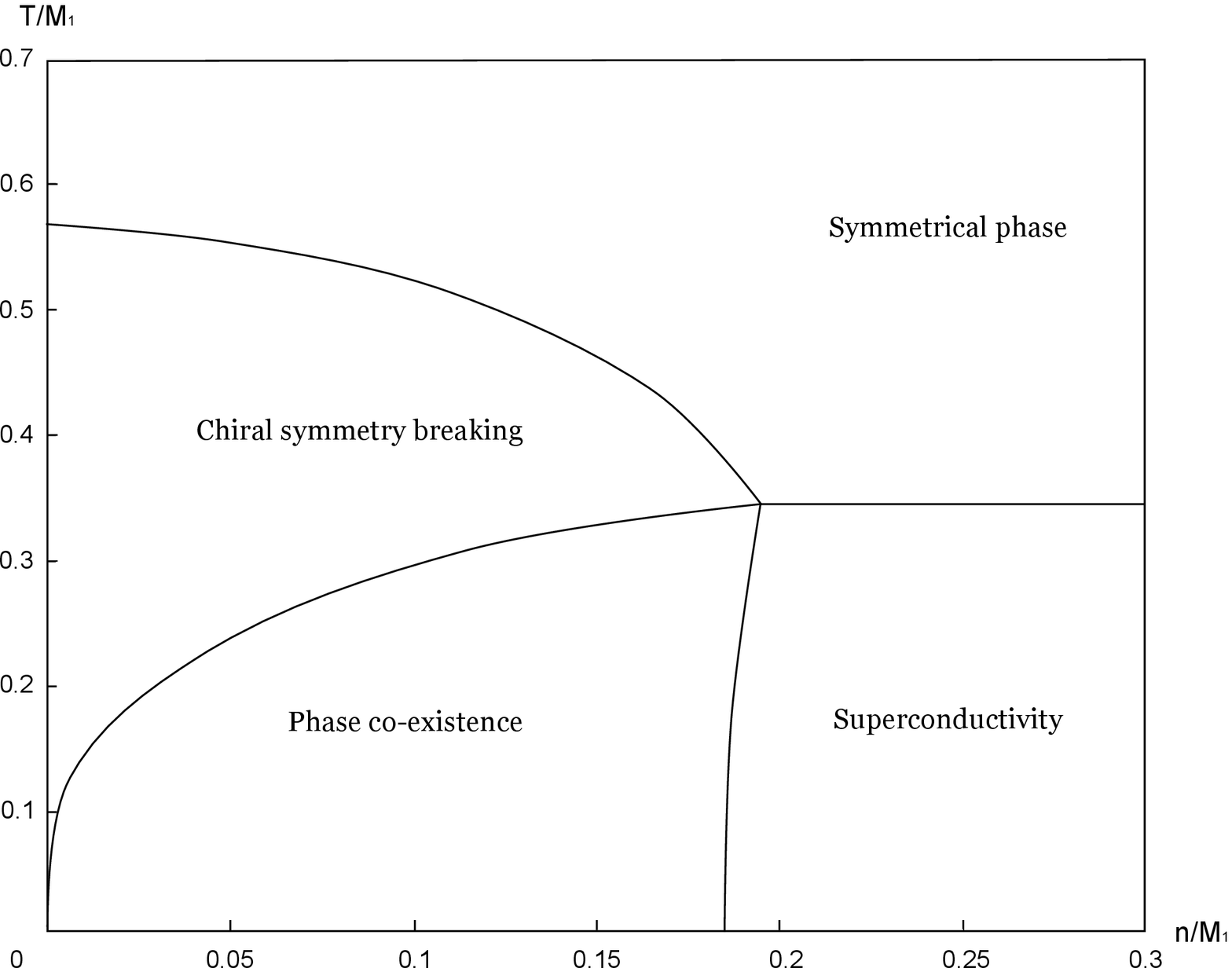}
\hfill
\includegraphics[width=0.5\textwidth]{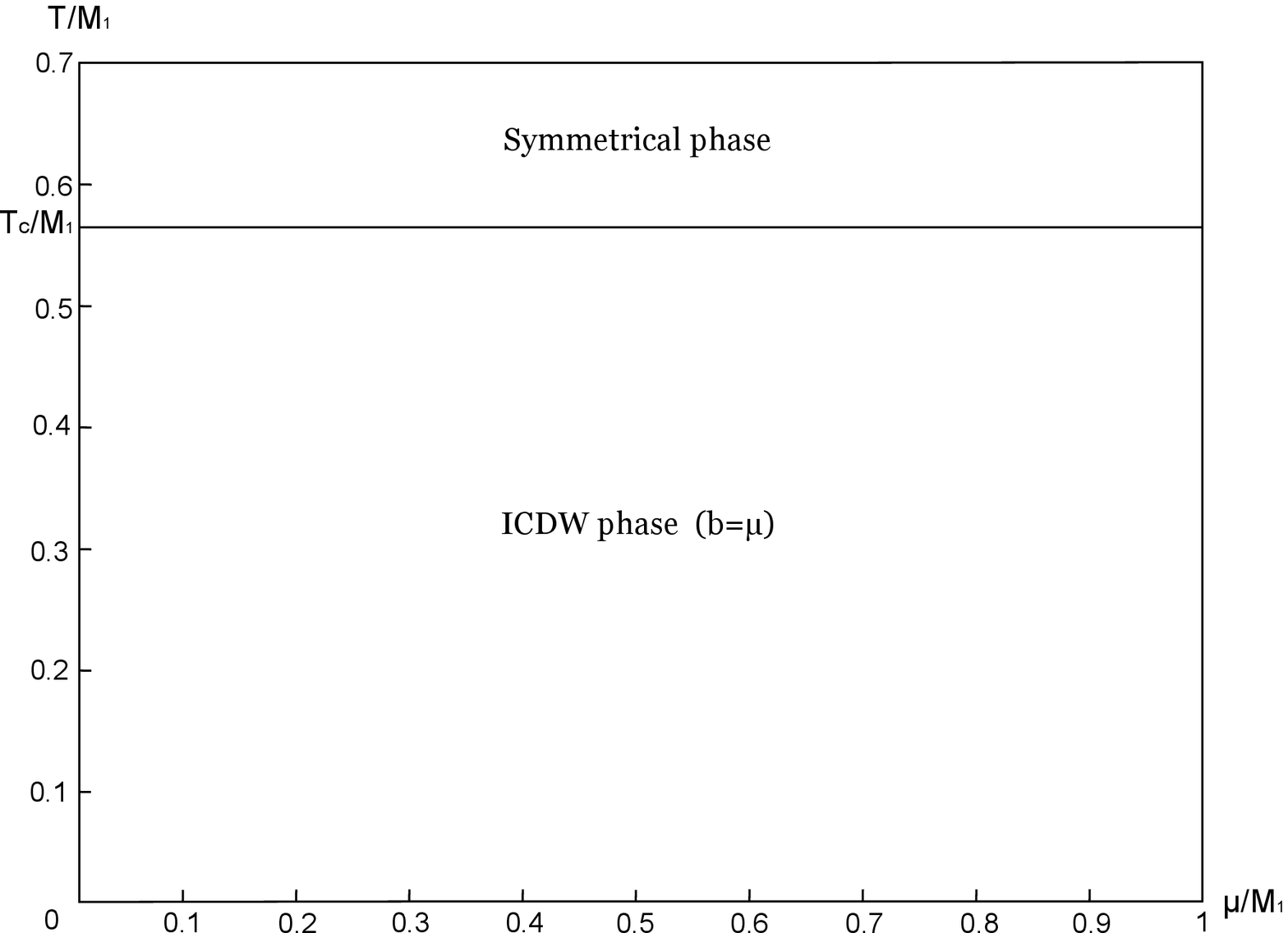}
\\
\parbox[t]{0.45\textwidth}{
 \caption{Phase structure of the model in terms of particle density
   $n$ and temperature $T$ at $\delta=1$ in the homogeneous case of
   the ansatz (\ref{8}) for condensates $(b=b'=0)$. } } \hfill
\parbox[t]{0.5\textwidth}{
 \caption{$(\mu,T)$-phase structure of the model at $\delta=1$, when
   the possibility for condensate inhomogeneity in the framework of
   the  ansatz (\ref{8}) is taken into account. At $T<T_c\approx
   0.57M_1$ the inhomogeneous chiral density wave (ICDW) phase is
   arranged, where we have for condensates:
 $\Delta=0$, $b'=0$, $M\ge 0$, and $b=\mu$. In the symmetrical phase  $\Delta=b'=M=b=0$. }}
\end{figure}

\section{Inhomogeneous case of the ansatz (\ref{8}) for condensates: $b\neq
  0$, $b'\neq 0$}

In this section we determine all quantities entering the ansatz
(\ref{8}) for condensates dynamically, i.e. we search for $\Delta$,
$M$, $b$, and $b'$ taking into account the stationarity equations for
the TDP (\ref{11}). In particular, it means that from the very
beginning the quantities $b$ and $b'$ are not supposed to be zero a
priory (as in the previous section). In this case the TDP (\ref{11})
of the system at $T=0$ can be transformed to the following expression:
\begin{eqnarray}
\Omega (M,b,b',\Delta)&=&
\frac{M^2}{4G_1}+\frac{\Delta^2}{4G_2}
+\frac{i}{2}\int\frac{d^2p}{(2\pi)^2}\ln\Big [\Delta^4-2\Delta^2(p_0^2-p_1^2+M^2+b'^2-\tilde\mu^2)\nonumber\\
&+&\big (M^2+(p_1-b')^2-(p_0-\tilde\mu)^2\big )\big (M^2+(p_1+b')^2-
(p_0+\tilde\mu)^2\big )\Big ], \label{25}
\end{eqnarray}
where $\tilde\mu=\mu-b$. The expression (\ref{25}) of the TDP
resembles the thermodynamic potential of the (1+1)--dimensional
model with inhomogeneous charged pion condensation phenomenon (see,
e.g., the expression (17) in \cite{gubina2}, where the corresponding
TDP is presented). Thus, using the same rather tedious technique as
in \cite{gubina2}, it is possible to show that at arbitrary fixed
$M,b,\Delta$ the absolute minimum of the function (\ref{25}) vs $b'$
always occurs at $b'=0$. This means that in the framework of the
ansatz (\ref{8}), the diquark condensate is always  spatially
homogeneous within the model (1). Now, taking into account this
result, we suppose that $b'=0$ in what follows.
\footnote{To avoid overloading of our paper by extra formulas, we omit a detailed proof of this fact. Moreover, earlier in the paper
\cite{ohwa} the same result was obtained in the particular case of
the model (1) with $G_1=0$.} As a consequence, in this case the
unrenormalized TDP (\ref{11}) looks like the expressions
(\ref{12})-(\ref{14}) in which $\mu$ should be replaced by
$\tilde\mu\equiv\mu-b$, i.e.
\begin{eqnarray}
\Omega^{un} (M,b,\Delta)\equiv \Omega^{un} (M,b,b',\Delta)\Big |_{b'=0}&=&
\frac{M^2}{4G_1}+\frac{\Delta^2}{4G_2}-\frac{1}{2\pi}
\int_{0}^\infty dp_1\tilde E_+-\frac{1}{2\pi}\int_{0}^\infty dp_1\tilde E_-, \label{27}
\end{eqnarray}
where
\begin{eqnarray}
\tilde E_\pm^2=\tilde\mu^2+p_1^2+M^2+\Delta^2\pm 2\sqrt{M^2\Delta^2+\tilde\mu^2(p_1^2+M^2)}.
 \label{28}
\end{eqnarray}
In contrast to the TDP (\ref{14}) of the previous section,
the TDP (\ref{27}) is a function of three variables, $M$, $\Delta$ and
$b$. Since, in this case $b$ simply shifts effectively the chemical
potential $\mu$, one can consider the TDP (\ref{27}) as a function of
$M$, $\Delta$ and  $\tilde\mu=\mu-b$. Clearly, without loss of
generality one can study the expression (\ref{27}) in the region $M\ge
0$, $\Delta\ge 0$ and $\tilde\mu\ge 0$.

To find a finite renormalized expression for the TDP (\ref{27}), we
should first regularize it and then perform a renormalization
procedure in order to remove at $\Lambda\to\infty$ the UV
divergences by demanding an appropriate behavior of the bare
coupling constants $G_{1,2}$ vs the cutoff parameter $\Lambda$. In
the case of spatially homogeneous condensates all regularization
schemes are usually equivalent. However, in the case of spatially
inhomogeneous condensates the translational invariance over one or
several spatial coordinates is lost. So, the corresponding (spatial)
momenta are not conserved. Then, if one uses the momentum-cutoff
regularization technique, as in the previous section, nonphysical
(spurious) $b$-dependent terms appear, and the TDP acquires some
non-physical properties such as unboundedness from below with
respect to $b$, etc. In order to obtain a physically reliable TDP
(or effective potential), in this case an additional substraction
procedure is usually applied (for details see in
\cite{miransky,gubina}). On the other hand, if one uses more
adequate regularization schemes such as Schwinger proper-time
\cite{nakano,nickel} or energy-cutoff regularizations
\cite{gubina2,zfk}, etc., such spurious terms do not appear.
\footnote{As discussed in the recent papers
\cite{gubina,nakano,nickel,gubina2,zfk}, an adequate regularization
scheme in the case of spatially inhomogeneous phases consists in the
following: for different quasiparticles the same restriction on
their region of energy values $\tilde E_\pm$ should be used in a
regularized thermodynamic potential.}

In the present paper the slightly modified energy cutoff
regularization
scheme of \cite{zfk} is adopted. (See also \cite{gubina,ohwa}, where a
similar regularization was used in searching for both chiral density
waves and inhomogeneous diquark condensate in some NJL$_2$ models.)
Namely, we require that only
energies with momenta $p_1$, constrained by the
relations  $\tilde E_\pm(M=0,\Delta=0)=p_1\pm\tilde\mu<\Lambda$,
contribute to the regularized thermodynamic potential. This means
that the term with energy $\tilde E_+$ ($\tilde E_-$) in (\ref{27})
should be integrated in the regularized expression for TDP over the
interval $0<p_1<\Lambda-\tilde\mu$
($0<p_1<\Lambda+\tilde\mu$). Consequently, instead of (\ref{27})
we have the following regularized TDP:
\begin{eqnarray}
\Omega^{reg}(M,b,\Delta)&=&\frac{M^2}{4G_1}+\frac{\Delta^2}{4G_2}-\frac{1}{2\pi}
\int_{0}^{\Lambda-\tilde\mu} dp_1\tilde E_+-\frac{1}{2\pi}\int_{0}^{\Lambda+\tilde\mu} dp_1\tilde E_-, \label{34}
\end{eqnarray}
The expression (\ref{34}) can be presented in the following form:
\begin{eqnarray}
\Omega^{reg}(M,b,\Delta)&=&\widetilde\Omega^{reg}(M,b,\Delta)+\frac{1}{2\pi}
\int_{\Lambda-\tilde\mu}^\Lambda dp_1\tilde E_+-\frac{1}{2\pi}\int_{\Lambda}^{\Lambda+\tilde\mu} dp_1\tilde E_-, \label{35}
\end{eqnarray}
where $\widetilde\Omega^{reg}(M,b,\Delta)$ is the TDP (\ref{15}) with
simple replacement $\mu\to\tilde\mu$. Replacing in this formula
$G_{1,2}$ by $G_{1,2}(\Lambda)$ from (\ref{18}) and adding an
unessential constant
$\Lambda^2/2\pi$, we obtain in the limit $\Lambda\to \infty$ the
renormalized expression
\begin{eqnarray}
\Omega(M,b,\Delta)=\widetilde\Omega(M,b,\Delta)+\frac{\tilde\mu^2}{2\pi}-\frac{\mu^2}{2\pi}, \label{36}
\end{eqnarray}
where
\begin{eqnarray}
\widetilde\Omega (M,b,\Delta)&=&V_0 (M,\Delta)
-\int_{0}^\infty\frac{dp_1}{2\pi}\Big (\tilde E_++\tilde E_--\sqrt{p_1^2+(M+\Delta)^2}-\sqrt{p_1^2+(M-\Delta)^2}\Big ) \label{29}
\end{eqnarray}
and $V_0 (M,\Delta)$ is given in (\ref{22}). Note that the TDP
(\ref{29}) is formally equal to the renormalized TDP (\ref{24}) of the
homogeneous condensates case, in which one should simply perform the
replacement $\mu\to\tilde\mu$. To obtain the last terms in (\ref{36}),
one should take into
account that at $\Lambda\to\infty$ the $p_1$-values in both
integrals in (\ref{35}) are much greater than $M,\Delta,b,\mu$. In
this case,
it is possible to expand the quantities $\tilde E_\pm$ into power
series of $p_1$ and then to integrate each term. Moreover, we add in
the expression (\ref{36}) an unessential $b$-independent term,
-$\mu^2/2\pi$, in order to reproduce at $b=0$ the TDP (\ref{24}),
corresponding to a spatially homogeneous chiral condensate.

{\bf Phase structure at $T=0$.} It is clear that to find the phase
portrait of the model at $T=0$,  one should investigate the global
minimum point (GMP) of the TDP $\Omega(M,b,\Delta)$ (\ref{36}) vs
the dynamical variables $M,b,\Delta$. Since in our case the variable
$b$ is absorbed by the chemical potential, the TDP (\ref{36}) is
indeed a function of three variables
$M,\Delta,\tilde\mu\equiv\mu-b$. Thus,  searching for the GMP of
this function consists effectively of two stages. First, one can
find the extremum of this function over $M$ and $\Delta$ (taking
into account the results of  section \ref{mu}) \footnote{As in the
case with $b=0$, in the inhomogeneous case we did not find local
minima of the TDP (\ref{36}) in which both $M\ne 0$ and $\Delta\ne
0$. } and then one minimizes the obtained expression over the
variable $\tilde\mu$. Following this strategy, let us introduce the
quantity
\begin{eqnarray}
\Omega(\tilde\mu)=\operatornamewithlimits{min}_{M\ge 0,\Delta\ge 0}\Big\{\Omega (M,b,\Delta)\Big\}.
\label{40}
\end{eqnarray}
Now suppose that $\delta>0$. Taking into account the results of the
investigation of the
GMP of the TDP (\ref{24}) (see subsection \ref{mu}), it is easy to see
that at $\tilde\mu<\mu_c$ with $\mu_c$ given in (\ref{39}), the
function $\Omega(M,b,\Delta)$ is minimized at $M=M_1$, $\Delta=0$
(recall, $\tilde\mu$ is fixed). In this case
$4\pi\Omega(\tilde\mu)=-M_1^2-2\mu^2+2\tilde\mu^2$. However, at
$\tilde\mu>\mu_c$ the minimum of $\Omega(M,b,\Delta)$ is reached at
$M=0,\Delta=M_1\exp (-\delta/2)$. Hence, in this case
$4\pi\Omega(\tilde\mu)=-M_1^2\exp (-\delta)-2\mu^2$. Therefore, the
function (\ref{40}) takes the form
\begin{eqnarray}
4\pi\Omega(\tilde\mu)=\left\{\begin{array}{ll} -M_1^2-2\mu^2+2\tilde\mu^2,~~ & if~~~~ \tilde\mu<\mu_c;\\
-M_1^2\exp (-\delta)-2\mu^2,~~ & if ~~~~\tilde\mu>\mu_c.\end{array}\right.
\label{41}
\end{eqnarray}
Since for arbitrary fixed $\mu>0$ the minimum of the function
(\ref{41})  over the variable $\tilde\mu\ge 0$ occurs at
$\tilde\mu=0$, i.e. at $b=\mu$, we conclude that for arbitrary
values of $\delta>0$ and $\mu>0$ the spatially inhomogeneous phase
in the form of chiral spirals (chiral density waves) is more
preferable than either of the two homogeneous phases, homogeneous
chiral symmetry breaking phase or homogeneous phase with nonzero
diquark condensate.

In the case of $\delta<0$ the GMP of the TDP $\Omega(M,b,\Delta)$
over two  variables, $M$ and $\Delta$, lies at the point
$(M=0,\Delta=M_1\exp (-\delta/2))$ (see subsection \ref{mu}). Hence,
in this case $4\pi\Omega(\tilde\mu)=-M_1^2\exp (-\delta)-2\mu^2$,
i.e. it does not depend on $\tilde\mu$. As a result, we could take
$b=0$ which corresponds to the homogeneous diquark condensation
phase at arbitrary $\delta<0$ and $\mu>0$ values, i.e. the
(homogeneous)
superconductivity phenomenon can appear in the model only at
$G_2>G_1$. In the case $G_2<G_1$, it is suppressed by inhomogeneous
chiral density wave phase. In contrast, if the homogeneous ansatz
for condensates is used (see \cite{chodos}, as well as the previous
section), then Cooper pairing appears in the model at arbitrary
$G_2<G_1$ starting from $\mu>\mu_c$ (\ref{39}).

{\bf Phase structure at $T\ne 0$.} In order to include the
temperature into consideration, let us start from the unrenormalized
expression (\ref{11}) for the TDP. Then at $b'=0$ we have
\begin{eqnarray}
\Omega^{un} (M,b,\Delta)= \frac{M^2}{4G_1}+\frac{\Delta^2}{4G_2}
+\frac{i}{2}\int\frac{d^2p}{(2\pi)^2}\ln\Big [(p_0^2-\tilde
E_+^2)(p_0^2-\tilde E_-^2)\Big ].  \label{42}
\end{eqnarray}
(Obviously, the TDP (\ref{42}) is equal to the expression
(\ref{27}).) To get the corresponding unrenormalized thermodynamic
potential $\Omega^{un}_{\scriptscriptstyle{T}}(M,b,\Delta)$ in the
case of nonzero temperature, one again performs the standard
replacements (\ref{430}) in (\ref{42}). Summing over Matsubara
frequencies in the obtained expression, one finds
\begin{eqnarray}
\Omega^{un}_{\scriptscriptstyle{T}}(M,b,\Delta)\!
&=&\frac{M^2}{4G_1}+\frac{\Delta^2}{4G_2}-\int_{0}^{\infty}\frac{dp_1}{2\pi}
\Big\{\tilde E_++\tilde E_-\Big\}-T\int_{0}^{\infty}\frac{dp_1}{\pi}\ln\big\{ [1+e^{-\beta
\tilde E_+}][1+e^{-\beta
\tilde E_-}]\big\}\nonumber\\&=&\Omega^{un}(M,b,\Delta)-T\int_{0}^{\infty}\frac{dp_1}{\pi}\ln\big\{ [1+e^{-\beta
\tilde E_+}][1+e^{-\beta
\tilde E_-}]\big\}, \label{1202}
\end{eqnarray}
where $\beta=1/T$, $\tilde E_\pm$ are given in (\ref{28}), and
$\Omega^{un}(M,b,\Delta)$ is the unrenormalized TDP (\ref{27}) at
$T=0$. Since the last integral in (\ref{1202}) is convergent, we
should again renormalize the term $\Omega^{un}(M,b,\Delta)$ using
the same regularization scheme as above in this section (see
expression (\ref{34}) for the corresponding regularized TDP), in
order to obtain a finite renormalized TDP
$\Omega_{\scriptscriptstyle{T}}(M,b,\Delta)$ at nonzero temperature.
As a result, we have
\begin{eqnarray}
\Omega_{\scriptscriptstyle{T}}(M,b,\Delta)\!
&=&\Omega(M,b,\Delta)-T\int_{0}^{\infty}\frac{dp_1}{\pi}\ln\big\{
[1+e^{-\beta \tilde E_+}][1+e^{-\beta \tilde E_-}]\big\},
\label{43}
\end{eqnarray}
where $\Omega(M,b,\Delta)$ is given in (\ref{36}). After numerical
investigations of this TDP we obtain, e.g., at $\delta=1$ the
$(\mu,T)$-phase portrait in Fig. 5. It is clear from this figure
that at rather low temperatures $T<T_c$ and arbitrary values of
chemical potential, the chiral symmetry breaking phase with
inhomogeneous condensate in the form of chiral spiral is arranged
(in Fig. 5 it is denoted as inhomogeneous chiral density wave (ICDW)
phase). In this phase $\Delta=0$, $M\ne 0$, and wave vector $b$ does
not depend on $T$ and equals $\mu$ (at each fixed value of $\mu$ the
gap $M$ decreases in  this phase continuously from $M_1$ to zero,
when temperature varies from zero to $T_c$).

If $\delta<0$, then in the framework of the ansatz (\ref{8}) we have
not found any phase with nonzero values of $b$ and/or $b'$, so in
this case the $(\mu,T)$-phase portrait of the model is the same as in
Fig. 2.

Finally, a few words about the particle density behavior as well as
the density-temperature phase portrait of the model, when  spatial
inhomogeneity of condensates in the form of the ansatz (\ref{8}) is
allowed. In this case we denote the particle density of the model by
$\tilde n$,
\begin{eqnarray}
\tilde n=-\partial\Omega_{\scriptscriptstyle{T}} (M_0,b_0,\Delta_0)/\partial\mu,
 \label{01203}
\end{eqnarray}
where the TDP $\Omega_{\scriptscriptstyle{T}}(M,b,\Delta)$ is given in
(\ref{43}) and $M_0,b_0,\Delta_0$ are the coordinates of its global
minimum point. It is evident that in the symmetrical phase of the
model, where $\Delta_0=M_0=b_0=0$, the particle density (\ref{01203})
is equal to  expression (\ref{1204}), i.e. $\tilde
n_{\scriptscriptstyle{SYM}}=\mu/\pi$.

In contrast, in the ICDW phase (see Fig. 5) we have $b_0=\mu,
\Delta_0=0$, and $M_0\ne 0$ (the last quantity is a $T$-dependent
one). So at $b=\mu$ the integral term of the expression (\ref{43})
does not depend on $\mu$ and, hence, does not contribute to the
particle density (\ref{01203}) in the ICDW phase. Moreover, as it
follows from the expressions (\ref{36}) and (\ref{29}), in this case,
i.e. at $b=\mu$, the TDP $\Omega(M,b,\Delta)$ in (\ref{43}) can be
presented in the form: $\Omega(M,b=\mu,\Delta)=-\mu^2/(2\pi)+\cdots$,
where we have omitted  all the $\mu$-independent terms. As a result,
it is easy to see from (\ref{01203}) that inside of the ICDW phase the
particle density ($\equiv$ $\tilde n_{\scriptscriptstyle{ICDW}}$) does
not depend on $T$ and, furthermore, $\tilde
n_{\scriptscriptstyle{ICDW}}=\mu/\pi$. It clear from this discussion
that the particle density--temperature phase diagram of the model has a
rather trivial form, when the inhomogeneity ansatz (\ref{8}) for
condensates is applied. Indeed, one should take the $(\mu,T)$-phase
portrait (it is Fig. 5 at $\delta>0$ or Fig. 2 at $\delta<0$) and then
simply rename the horizontal $\mu$-axis in favor of $\tilde n$-axis.

It is clear from the above consideration that in the case of
inhomogeneous condensates the critical temperature, at which there
occurs a symmetry restoring phase transition, does not depend on
$\mu$ and/or $\tilde n$. However, it is a model dependent effect.
Indeed, in this paper we deal with the model (1) which is invariant
with respect to $U_A(1)$-chiral symmetry. In contrast, in
(1+1)-dimensional model with $SU_A(2)$-chiral symmetry \cite{gubina}
the temperature of a transition between chiral density wave- and
symmetrical phases is a $\mu$-dependent quantity. Moreover, in the
last model the phase diagram has a more complicated form (see
\cite{gubina} for details).

\section{CONCLUSIONS}

We have investigated the phase structure of the NJL$_2$-type model
(1) in the framework of the Fulde--Ferrel single plane wave
spatially non-uniform ansatz (\ref{8}) both for chiral and
superconducting condensates. The following results are obtained:

1) It is shown that spatially inhomogeneous superconducting
condensation is forbidden in the model (1) (of course, in the
framework of the ansatz (\ref{8})).

2) It is clear from Fig. 5 that in the case $G_1>G_2$ the
inhomogeneous chiral condensate in the form of the so-called chiral
density wave suppresses the appearance of the superconductivity at
arbitrary values of $\mu>0$. In contrast, if chiral- and Cooper
condensates are assumed to be  spatially uniform, then at
sufficiently high $\mu$ the superconducting phase is allowed to
exist in this case (see Fig. 1  and \cite{chodos}).

3) If spatially inhomogeneous ansatz (\ref{8}) for condensates is
taken into account in the model (1), then Cooper pairing (which is
spatially homogeneous) is possible only at sufficiently strong
interaction in the quark-quark channel, i.e. at $G_2>G_1$ (see Fig.
2).


\begin{thebibliography}{999}

\bibitem{3+1}
D.V. Deryagin, D.Y. Grigoriev and V.A. Rubakov, Int. J. Mod. Phys. A
{\bf 7}, 659 (1992); M.~Sadzikowski and W.~Broniowski, Phys.\ Lett.\
B {\bf 488}, 63 (2000);
 W.~Broniowski,
  Acta Phys.\ Polon.\ Supp.\  {\bf 5}, 631 (2012).

\bibitem{nakano}
 E.~Nakano and T.~Tatsumi,
  Phys.\ Rev.\  D {\bf 71}, 114006 (2005).

\bibitem{nickel}
D.~Nickel,  Phys.\ Rev.\  D {\bf 80}, 074025 (2009);
S.~Carignano, D.~Nickel and M.~Buballa,
 Phys.\ Rev.\  D {\bf 82}, 054009 (2010);
 H.~Abuki, D.~Ishibashi and K.~Suzuki,
  arXiv:1109.1615.

\bibitem{maedan}
 S.~Maedan, Prog.\ Theor.\ Phys.\  {\bf 123}, 285 (2010);
 A.~Flachi,
  JHEP {\bf 1201}, 023 (2012);
 arXiv:1304.6880 [hep-th].

\bibitem{Heinz:2013eu}
  A.~Heinz,
  arXiv:1301.3430 [hep-ph].

\bibitem{pisarski}
 T.~Kojo, Y.~Hidaka, L.~McLerran and R.D.~Pisarski,
  Nucl.\ Phys.\  A {\bf 843}, 37 (2010).

\bibitem{miransky}
 E.V.~Gorbar, M.~Hashimoto and V.A.~Miransky,
  Phys.\ Rev.\ Lett.\  {\bf 96}, 022005 (2006);
J.O.~Andersen and T.~Brauner,
  Phys.\ Rev.\  D {\bf 81}, 096004 (2010);
C.f.~Mu, L.y.~He and Y.x.~Liu,
  Phys.\ Rev.\  D {\bf 82}, 056006 (2010).

\bibitem{zfk}
I.E. Frolov, K.G. Klimenko and V.Ch. Zhukovsky,   Phys.\ Rev.\  D
{\bf 82}, 076002 (2010);
 Moscow Univ.\ Phys.\ Bull.\  {\bf 65}, 539 (2010).

\bibitem{incera}
E.J.~Ferrer, V.~de la Incera and A.~Sanchez,
  arXiv:1205.4492.

\bibitem{thies}
V.~Schon and M.~Thies, 
 Phys.\ Rev.\  D {\bf 62}, 096002 (2000);
A.~Brzoska and M.~Thies,
  Phys.\ Rev.\  D {\bf 65}, 125001 (2002).

\bibitem{thies2}
O.~Schnetz, M.~Thies and K.~Urlichs,
  Annals Phys.\  {\bf 314}, 425 (2004);
G.~Basar, G.V.~Dunne and M.~Thies,
  Phys.\ Rev.\  D {\bf 79}, 105012 (2009);
C.~Boehmer and M.~Thies,
   Phys.\ Rev.\  D {\bf 80}, 125038 (2009);
J.~Hofmann,  Phys.\ Rev.\  D {\bf 82}, 125027 (2010).

\bibitem{misha}
F.~Correa, G.V.~Dunne and M.S.~Plyushchay,
  Annals Phys.\  {\bf 324}, 2522 (2009).

\bibitem{gubina}
  D.~Ebert, N.V.~Gubina, K.G.~Klimenko, S.G.~Kurbanov, V.C.~Zhukovsky,
  Phys.\ Rev.\  {\bf D84}, 025004 (2011).

\bibitem{gubina2}
  N.V.~Gubina, K.G.~Klimenko, S.G.~Kurbanov and V.C.~Zhukovsky,
  Phys.\ Rev.\ D {\bf 86}, 085011 (2012).

\bibitem{gubina3}
  N.V.~Gubina, K.G.~Klimenko, S.G.~Kurbanov and V.C.~Zhukovsky,
Moscow Univ.\ Phys.\ Bull.\  {\bf 67}, 131 (2012);
Yad. Fiz. {\bf 76}, 1443 (2013).

\bibitem{Caldas}
H.~Caldas and M.~A.~Continentino,
  J.\  Phys.\  B: At.\  Mol.\  Opt.\  Phys.\  {\bf 46}, 155301 (2013);
D. Roscher, J. Braun and J.E. Drut,
    arXiv:1311.0179  [cond-mat.quant-gas].

\bibitem{chodos}
 A.~Chodos, H.~Minakata, F.~Cooper, A.~Singh, and W.~Mao,
  Phys. Rev. D {\bf 61}, 045011 (2000).

\bibitem{ff}
P. Fulde and R.A. Ferrel, Phys. Rev. {\bf 135}, A550 (1964).

\bibitem{fujikawa}
K. Fujikawa,   Phys.\ Rev.\  D {\bf 21}, 2848 (1980).

\bibitem{kzz}
  K.G.~Klimenko, R.N.~Zhokhov and V.C.~Zhukovsky,
  Phys.\ Rev.\ D {\bf 86}, 105010 (2012).

\bibitem{Ebert:2009ty}
  D.~Ebert, K.G.~Klimenko,  Phys.\ Rev.\  {\bf D80}, 125013 (2009).

\bibitem{jacobs}
L. Jacobs, Phys.\ Rev.\  D {\bf 10}, 3956 (1974);
K.G.~Klimenko, Theor.\ Math.\ Phys.\  {\bf 70}, 87 (1987).

\bibitem{ohwa}
K.~Ohwa, Phys.\ Rev.\  D {\bf 65}, 085040 (2002).

\bibitem{Ebert:2010eq}
  D.~Ebert and K.G.~Klimenko,
  Phys.\ Rev.\ D {\bf 82}, 025018 (2010).

\end{thebibliography}
\end{document}